\documentclass[fleqn,usenatbib]{mnras}

\usepackage{mathptmx}
\usepackage[T1]{fontenc}

\DeclareRobustCommand{\VAN}[3]{#2}
\let\VANthebibliography\thebibliography
\def\thebibliography{\DeclareRobustCommand{\VAN}[3]{##3}\VANthebibliography}

\usepackage{graphicx}	% Including figure files
\usepackage{amsmath}	% Advanced maths commands
\usepackage{amssymb}	% Extra maths symbols

\title[Solar-wind model inputs]{Using in-situ solar-wind observations to generate inner-boundary conditions to outer-heliosphere simulations, 1: Dynamic time warping applied to synthetic observations}

\author[M.J. Owens and J.D. Nichols]{
Mathew J. Owens,$^{1}$\thanks{E-mail:m.j.owens@reading.ac.uk}
Jonathan D. Nichols$^{2}$
\\
$^{1}$Department of Meteorology, University of Reading, Earley Gate, Reading RG6 6BB\\
$^{2}$Physics and Astronomy Department, University of Leicester, University Road, Leicester, LE1 7RH}

\date{Accepted XXX. Received YYY; in original form ZZZ}

\pubyear{2021}

\begin{document}
\label{firstpage}
\pagerange{\pageref{firstpage}--\pageref{lastpage}}
\maketitle

% Abstract of the paper
\begin{abstract}
The structure and dynamics of the magnetospheres of the outer planets, particularly Saturn and Jupiter, have been explored both through remote and in-situ observations. Interpreting these observations often necessitates simultaneous knowledge of the solar-wind conditions impinging on the magnetosphere. Without an available upstream monitor, solar-wind context is typically provided using models initiated with either the output of magnetogram-constrained coronal models or, more commonly, in-situ observations from 1 AU. While 1-AU observations provide a direct measure of solar wind conditions, they are single-point observations and thus require interpolation to provide inputs to outer-heliosphere solar-wind models. In this study we test the different interpolation methods using synthetic 1-AU observations of time-evolving solar-wind structure. The simplest method is ``corotation'', which assumes solar-wind structure is steady state and rotates with the Sun. This method of reconstruction produces discontinuities in the solar-wind inputs as new observations become available. This can be reduced by corotating both back and forward in time, but this still introduces large errors in the magnitude and timing of solar wind streams. We show how the dynamic time warping (DTW) algorithm can provide around an order-of-magnitude improvement in solar-wind inputs to outer-heliosphere model from in-situ observations near 1 AU. This is intended to build the foundation for further work demonstrating and validating methods to improve inner-boundary conditions to outer-heliosphere solar-wind models, including dealing with solar wind transients and quantifying the improvements at Saturn and Jupiter. 
\end{abstract}

% Select between one and six entries from the list of approved keywords.
% Don't make up new ones.
\begin{keywords}
solar wind -- planet–star interactions -- Sun:heliosphere 
\end{keywords}

%%%%%%%%%%%%%%%%%%%%%%%%%%%%%%%%%%%%%%%%%%%%%%%%%%

%%%%%%%%%%%%%%%%% BODY OF PAPER %%%%%%%%%%%%%%%%%%
%================================================================================
\section{Introduction}
%================================================================================
The magnetospheres of the Saturn \citep{dougherty_saturn_2009} and Jupiter \citep{bagenal_jupiter_2004} have been explored both through remote observations and with in-situ spacecraft observations  \citep[e.g. ][]{clarke_response_2009, crary_solar_2005, hess_solar_2012, connerney_jupiters_2017}. Though the dynamics of the outer planets' magnetospheres are rotationally-dominated, they are strongly modulated by the solar wind, such that, like at Earth, understanding the dynamics of these magnetospheric systems requires simultaneous knowledge of the local solar wind conditions \citep{delamere_solar_2010, kennel_is_1975, nichols_magnetopause_2006, brice_magnetospheres_1970, badman_significance_2007, nichols_response_2017, masters_dayside_2015}. In the absence of such available observations, this must instead be provided by solar wind models.

Solar wind conditions at Earth are routinely forecast on the basis of photospheric observations, through coupled coronal and heliospheric models  \citep[e.g.][]{riley_empirically-driven_2001,odstrcil_modeling_2003,toth_space_2005,merkin_time-dependent_2016,pomoell_euhforia_2018}. However, for outer-planet research, genuine forecasting is rarely required and the accuracy of the solar wind structure reconstructed by magnetogram-based methods may not be sufficient \citep[e.g. ][]{owens_metrics_2008}. Thus the preferred method is to propagate in-situ observations from 1 AU to the required planetary target. Over short radial distances, this can be done ballistically \citep{vennerstrom_magnetic_2003}. However, the solar wind undergoes a great deal of dynamic evolution between 1 AU and 5-10 AU \citep[e.g. ][]{hanlon_evolution_2004}, which we here refer to as the outer heliosphere. Thus propagation of the available observations is typically done with numerical fluid models, with magnetohydrodynamic (MHD) models used in both 1-dimension \citep{tao_magnetic_2005, zieger_statistical_2008} and 2-dimensions (e.g. the MSWIM2D model, based on the \citet{toth_space_2005} framework. See also \citet{keebler_mswim2d_2021}). This general approach supports a huge volume of planetary science \citep[e.g.][]{zarka_modulation_2007,kimura_long-term_2013, provan_planetary_2015, clarke_response_2009, nichols_variation_2009, kita_characteristics_2016}.

As 1-AU observations are rarely available in perfect radial alignment with the target planet, observations also have to be extrapolated in longitude. This is true for one-dimensional models \citep{tao_magnetic_2005,zieger_statistical_2008}, where either the model inputs or outputs can be extrapolated between the observation longitude and the target longitude, or for two-dimensional models \citep{keebler_mswim2d_2021} where all longitudes must be reconstructed from the available single-point observations to provide a full set of inner boundary conditions at 1 AU. In both cases, `corotation'' is used, wherein the conditions at a given Carrington longitude (i.e., in a frame rotating with the Sun) are assumed to be equal to previous or subsequent observations at the same Carrington longitude. Thus, the assumption is made of no time evolution.

The simplest form of corotation uses the previous observation at a required longitude, i.e.\ it looks only back in time, which enables it to also be used for forecasting \citep[e.g.][]{kohutova_improving_2016,thomas_evaluating_2018,turner_influence_2021}. When the solar wind structure (in the rotating frame of the Sun) is evolving with time, this produces sharp discontinuities in the reconstructed solar wind at longitudes where new observations are introduced. To mitigate this, corotation can be ``smoothed'' in time (and hence Carrington longitude) by interpolating between the previous and subsequent observations at a given longitude. This approach is used by the most-recent implementation of the two-dimensional MSWIM2D model \citep{keebler_mswim2d_2021}. 

In this study we produce synthetic in-situ observations from a model of time-evolving solar wind and test how well the corotation schemes can reconstruct the true state. It is shown that the time smoothing form of corotation produces an improvement over simple corotation, greatly reducing the artifical discontinuities, but that there are still significant errors in reconstructing a time-evolving solar wind structure. Ideally, the strengths of the magnetogram-constrained solar wind models would be combined with the information from in-situ observations using data assimilation \citep{lang_variational_2019, lang_improving_2021}. However, this is only currently possible with reduced-physics models, and is computationally intensive for the long runs required for outer-heliosphere simulations. While these techniques will become more accessible in the future, in this study we explore an alternative method for best exploiting in-situ observations as inner-boundary conditions to outer-heliosphere solar-wind simulations.

To explicitly allow for time evolution, we develop a reconstruction scheme based on dynamic time warping (DTW). This method is explained in Section \ref{sec:methods} and tested in Section \ref{sec:results}. This study is intended to be the first in a series examining how best to use 1-AU observations to provide inner boundary conditions to solar-wind models. Future papers will: use the inner boundary conditions generated in this study to drive solar wind simulations and quantify the effect on the the solar wind conditions reconstructed at the orbits of Saturn and Jupiter; outline and test a scheme for dealing with transient structures from interplanetary coronal mass ejections in the in-situ observations; and apply the methods to real observations from near-Earth and STEREO spacecraft.

%================================================================================
\section{Methods}
\label{sec:methods}
%================================================================================
\subsection{Synthetic Solar Wind Observations}

\begin{figure}
	\includegraphics[width=\columnwidth]{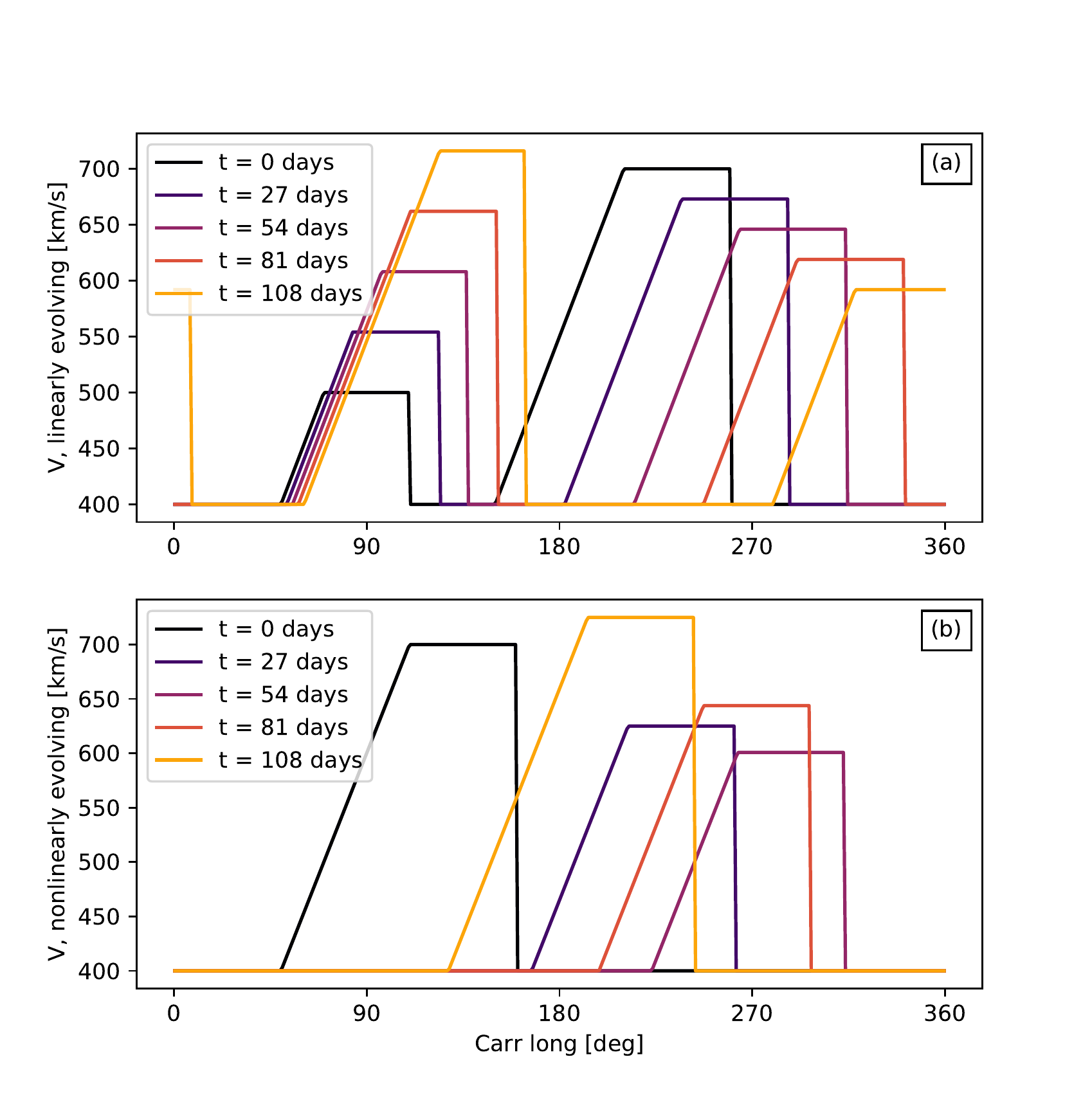}
    \caption{Two models of solar wind speed evolution. Snapshots of solar wind speed as a function of Carrington longitude are shown at five times, approximately one Carrington rotation apart. (a) Two fast solar wind streams evolve linearly in magnitude, width and position as a function of time.  (b) A single solar wind stream evolves nonlinearly in magnitude, width and position as a function of time.}
    \label{fig:models_lines}
\end{figure}

\begin{figure}
	\includegraphics[width=\columnwidth]{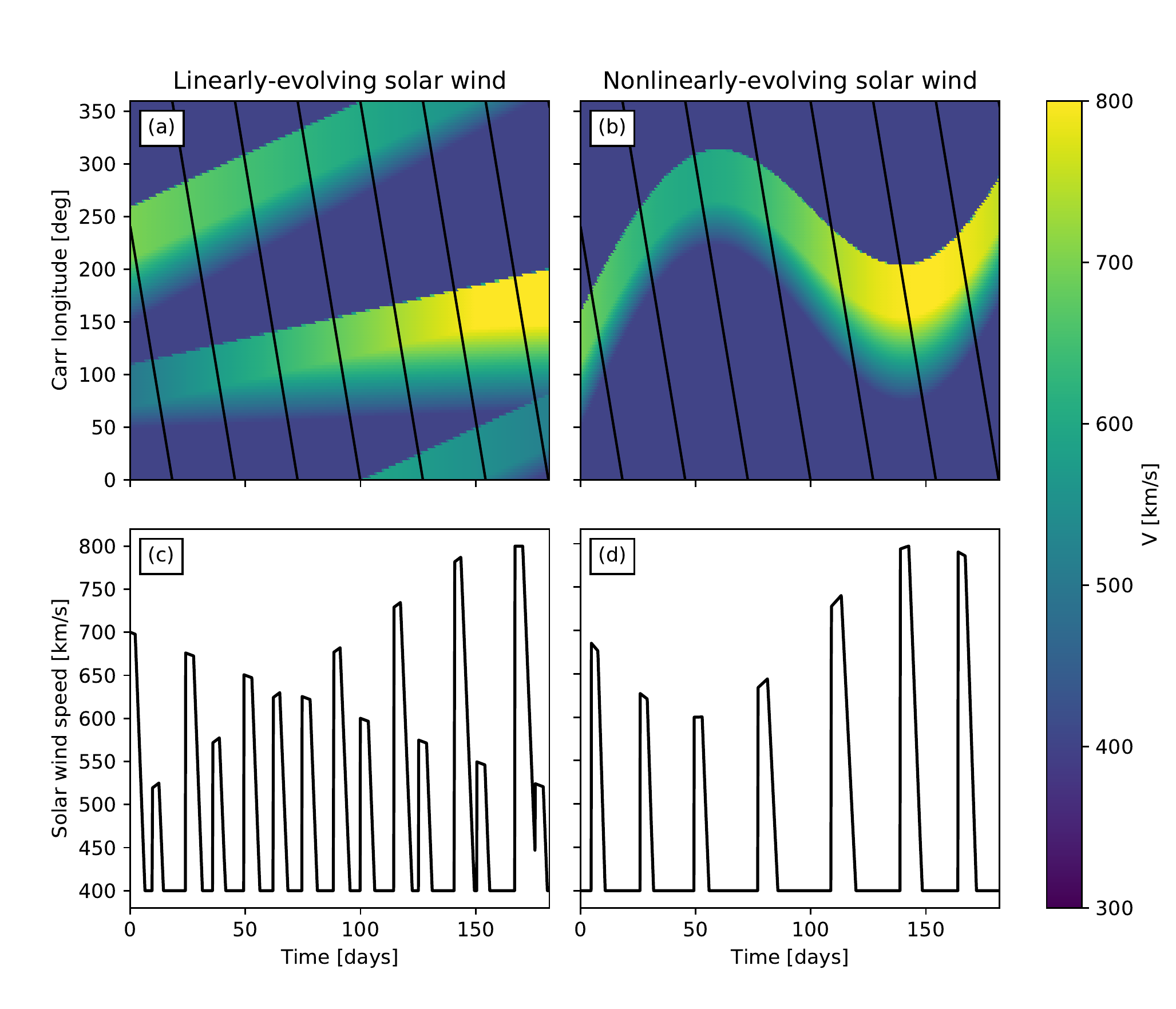}
    \caption{Two models of solar wind speed evolution. Left panels: Linearly-evolving $V$ with time. Right panels: Nonlinearly evolving $V$ with time. Top panels: Colour maps of solar wind speed as a function of Carrington longitude and time. Earth position is shown by the black lines. Bottom panels: The resulting $V$ time series in near-Earth space.}
    \label{fig:models_cmaps}
\end{figure}

In order to test solar wind reconstruction methods, we produce synthetic in-situ observations from time-evolving models of solar wind speed, $V$. This enables us to compare with the ``true'' state from which synthetic observations are constructed, which is not possible with real observations.

Figure \ref{fig:models_lines} shows two models of $V$ at the heliographic equator at 1 AU, in the Carrington frame rotating with the Sun. Figure \ref{fig:models_lines}a shows the linear model, in which two fast streams both move in longitude and change in magnitude linearly with time. The properties of the fast streams are chosen arbitrarily and can be varied in the analysis code available through the GitHub link in the acknowledgements. In the example used here, we choose values typical of the solar wind observed at 1 AU \citep[e.g.][]{owens_solar_2019}. Slow wind has a speed of 400 km s$^{-1}$. The first fast stream has an initial amplitude of 700 km s$^{-1}$ and begins at a maximum longitude of 260$^\circ$ with a width of $50^\circ$. The fast stream amplitude decays at a rate of 1 km s$^{-1}$ per day and it moves to greater Carrington longitude at a rate of $1^\circ$ per day. The second fast stream has an amplitude of 500 km s$^{-1}$, increasing at 2 km s$^{-1}$ per day, a width of 40$^\circ$, and an initial longitude of 110$^\circ$, increasing at $0.5^\circ$ per day. The maximum and minimum allowed speeds are capped at 800 km s$^{-1}$ and 400 km s$^{-1}$, respectively. Both fast streams have a declining $V$ back to the slow wind speed at a gradient of 5 km s$^{-1}$ per degree of longitude. As a (quasi-) stationary spacecraft at 1 AU moves to smaller Carrington longitude with time, these profiles produce $V$ time series of fast streams with a discontinuous rise in solar wind speed, followed by a constant peak $V$, before linearly declining back to slow wind. 

We also consider a nonlinear model, in which both the longitude and magnitude of the fast stream varies sinusoidally, shown by Figure \ref{fig:models_lines}b. In the example shown, there is slow wind of speed 400 km s$^{-1}$, with a single fast stream. The fast stream has a width of 50$^\circ$ and initially has a maximum longitude of 160$^\circ$. The longitude varies sinusoidally with an amplitude of 100$^\circ$ and a period of 200 days, as well as drifting linearly to greater longitude at a rate of 1 $^\circ$ per day. The amplitude of the fast stream is initially 600 km s$^\circ$, varying sinusoidally with an amplitude of 100 km s$^{-1}$ and a period of 200 days.

The $V$ obtained by these two models is shown by as a function of Carrington longitude and time in Figures \ref{fig:models_cmaps}a and \ref{fig:models_cmaps}b. In this format, a non-evolving solar wind would show horizontal bands of constant $V$. The black lines show Earth position, at which the synthetic $V$ time series is extracted and shown in Figures \ref{fig:models_cmaps}c and \ref{fig:models_cmaps}d. We next test the ability of various techniques to reconstruct $V$ at all Carrington longitudes from this synthetic near-Earth $V$ time series.

\subsection{Corotation Reconstruction Methods}

\begin{figure}
	\includegraphics[width=\columnwidth]{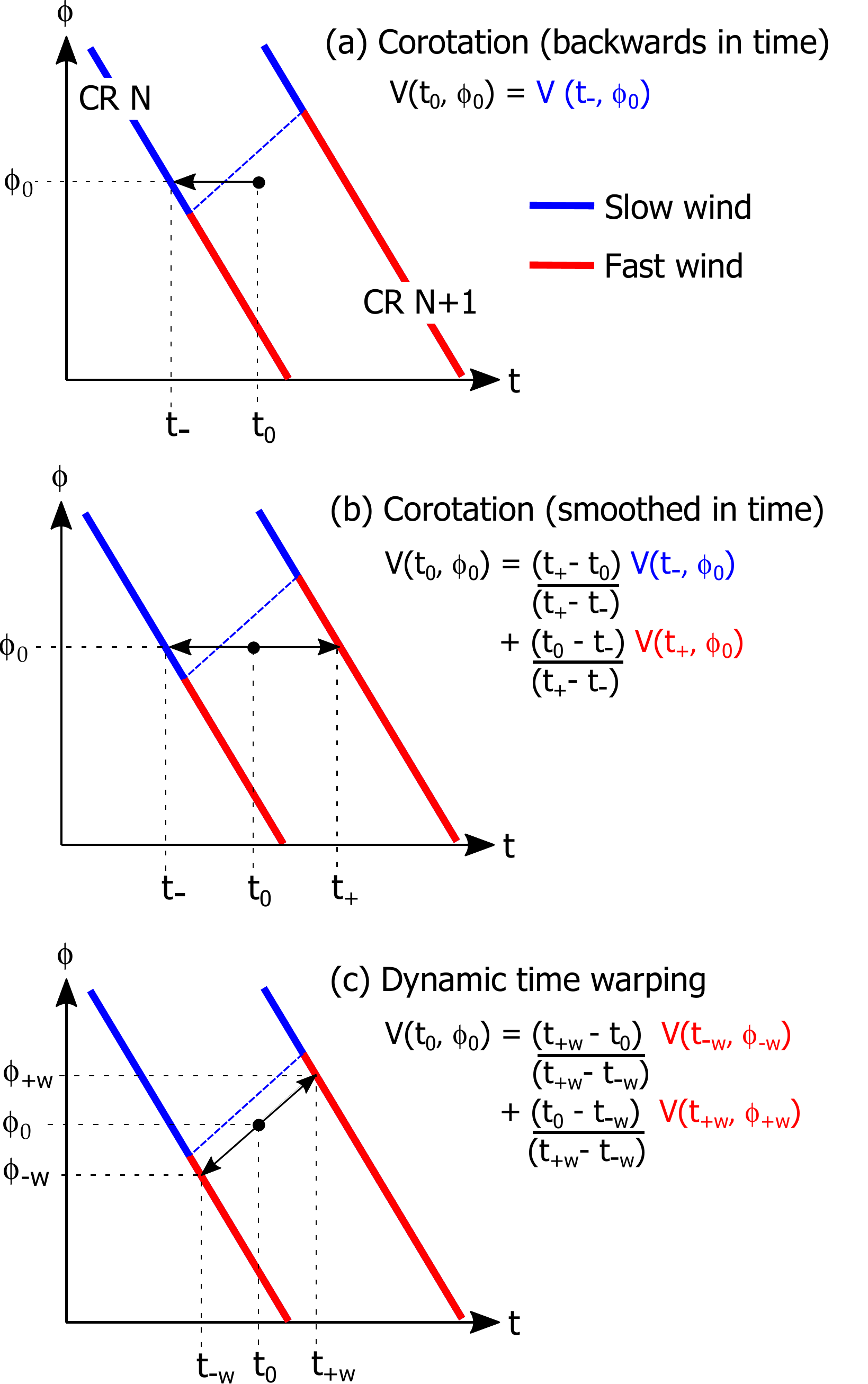}
    \caption{Schematics of the three reconstruction methods assessed in this study. Two Carrington rotations (CRs) of in situ observations are shown as a function of Carrington longitude, $\phi$, and time, $t$, with slow/fast wind in blue/red. The slow/fast wind interface (blue dashed line) moves to greater $\phi$ between CR $N$ to CR $N+1$. The reconstruction point, $(t_0,\phi_0)$, shown as a black dot, is in fast wind. (a) In the standard corotation method, $V(t_0,\phi_0)$ is equal to the value at the same Carrington longitude at the time of the previous observation. In this example, it gives slow wind. (b) Corotation smoothed in time uses a weighted average of the prior and subsequent values of $V$ at $\phi_0$, which results in intermediate speed wind. (c) Dynamic time warping aligns features in the CR $N$ and CR $N+1$ time series and takes a weighted average along the determined warp path. This results in fast wind at $V$, as required.}
    \label{fig:corotation}
\end{figure}

Cororation is the most commonly used technique to reconstruct solar wind conditions at longitudes distant to that where observations are available. Corotation assumes that the solar wind structure in the rotating frame of the Sun does not change with time. It is illustrated schematically in Figure \ref{fig:corotation}a. The thick diagonal lines track the position of Earth over two subsequent Carrington rotations, CR $N$ and $N+1$. The transition from red to blue shows an observed transition from slow to fast wind in near-Earth space. The slow-fast wind interface moves to larger Carrington longitude ($\phi$) with time, and thus the reconstruction point $(t_0, \phi_0)$ where the solar wind is to be reconstructed is in fast wind. Standard corotation looks back into the previous CR to find the previous $V$ observation at $\phi_0$, which occurred at a time $t_-$. Therefore the reconstructed $V$ at $(t_0, \phi_0)$ by corotation backwards in time, $V_{CB}$, is given by:

\begin{equation}
    V_{CB} (t_0, \phi_0) = V (t_-,\phi_0)
\end{equation}

In the example shown in Figure \ref{fig:corotation}a, the time evolution means that corotation (backwards in time) incorrectly estimates $V_{CB}(t_0, \phi_0)$ to be slow wind.  

For reconstruction (i.e. in non-forecast situations), corotation can be applied both forwards and backwards in time to find the prior and subsequent $V$ at the same Carrington longitude. A weighted average (based on timing proximity) of the two $V$ values is taken. We refer to this as `corotation time smoothing'. It is schematically shown in \ref{fig:corotation}b. The estimated speed by this method, $V_{CT}$, is given by:
\begin{equation}
    V_{CT} (t_0, \phi_0) = \frac{t_+ - t_0}{t_+ - t_-} V (t_-,\phi_0) +
    \frac{t_0 - t_-}{t_+ - t_-} V (t_+,\phi_0)
\end{equation}
where $t_+$ is the time of the of the $V$ observation at Carrington longitude $\phi_0$ during the subsequent Carrington rotation. 

In the example shown in Figure \ref{fig:corotation}b, the corotation time smoothing would produce intermediate speed solar wind at the reconstruction point.

\subsection{Reconstruction by Dynamic Time Warping}

\begin{figure}
	\includegraphics[width=\columnwidth]{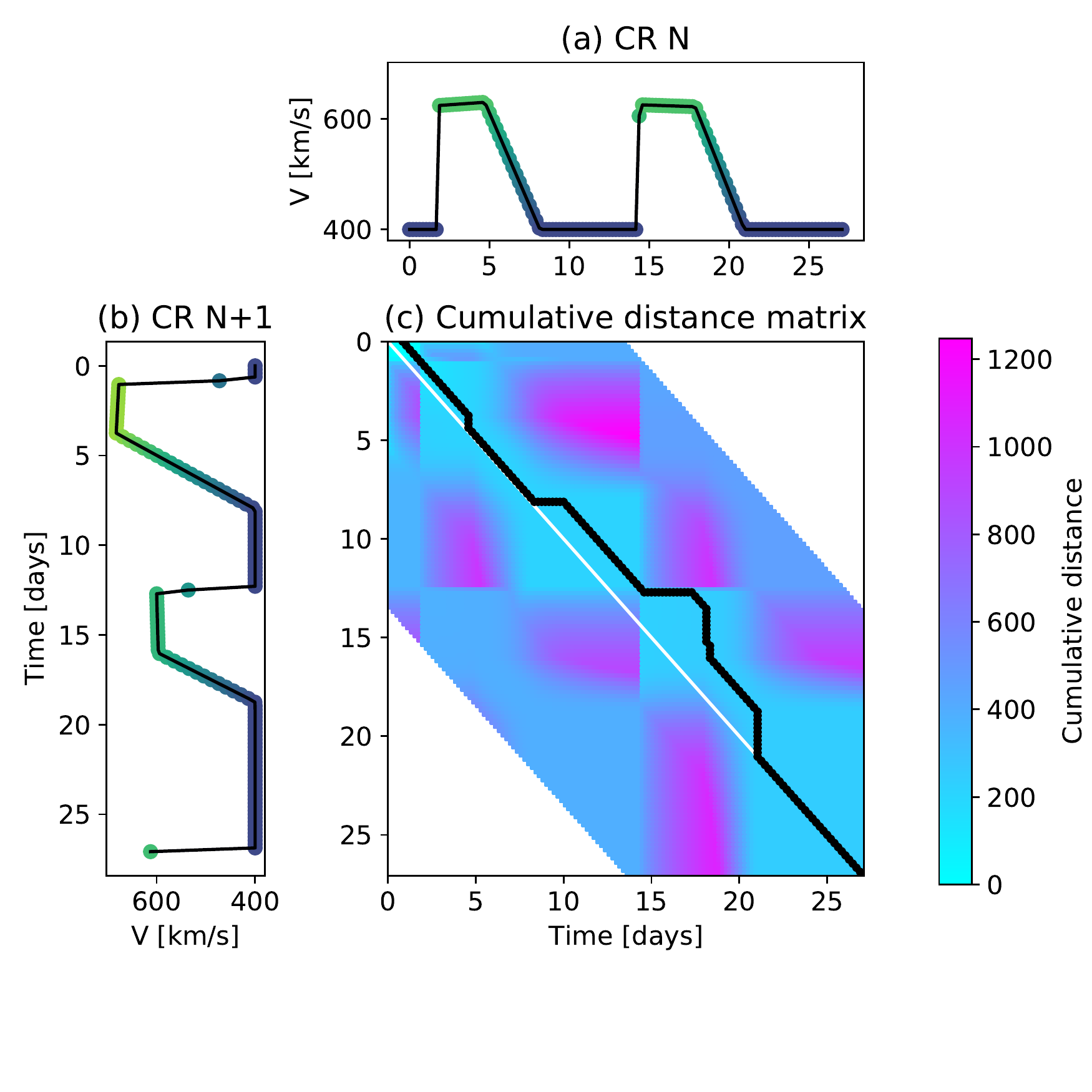}
    \caption{An example of dynamic time warping applied to two near-Earth $V$ time series produced by the linearly evolving model. Panel (a) shows the time series of $V$ for Carrington rotation $N$, panel (b) for Carrington rotation $N+1$. Fast/slow wind is further highlighted by yellow/blue. (c) A colour map of the total distance between time points for all possible connections. The black line traces the optimal path (i.e. minimum cumulative distance). For this example, most warp paths connect to earlier times in CR $N+1$, suggesting solar wind features are moving earlier in the Carrington rotation. The optimal path is best constrained at the leading edge of fast streams.}
    \label{fig:dtwmatrix}
\end{figure}

\begin{figure}
	\includegraphics[width=\columnwidth]{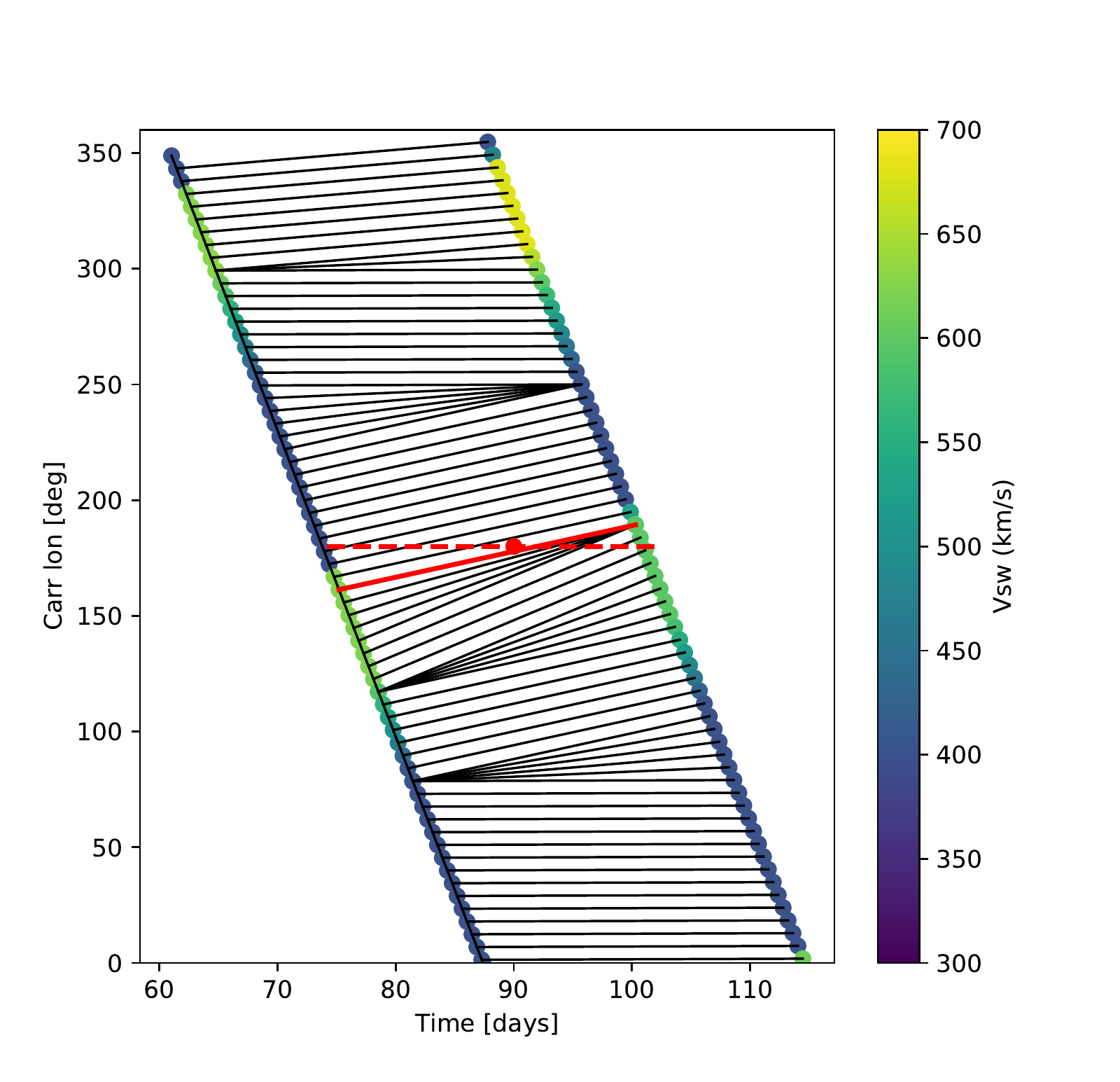}
    \caption{An example of dynamic time warping (DTW) reconstruction applied to subsequent Carrington rotations of near-Earth $V$ observations obtained from the linearly evolving $V$ model. The red dot shows the position/time where $V$ is to be reconstructed. Corotation uses previous/subsequent observed values in near-Earth space at the same Carrington longitude, shown by the dashed red line. The black lines show the DTW warp paths which best connect the $V$ time series at subsequent Carrington rotations. The DTW estimate of solar wind uses observed values along the closest warp path, shown by the solid red line.}
    \label{fig:dtw_explanation}
\end{figure}

As will be shown in Section \ref{sec:results},  corotation time smoothing provides improved reconstruction over simple back-in-time corotation, but it still does not explicitly account for time evolution of solar-wind structures. Here we match solar-wind structures in consecutive time series bounding the required reconstruction time/position to explicitly allow for time evolution. The principle is shown schematically in Figure \ref{fig:corotation}c. If features, such as the slow/fast solar-wind transition, can be connected between subsequent time series, then the time smoothing can be performed along ``connection lines'' rather than along constant Carrington longitude lines. In the example shown, this would correctly identify fast wind at the reconstruction point.

In order to systematically determine the connections between solar wind streams in two time series, we use an algorithm called `dynamic time warping' \citep[DTW, ][]{sakoe_dynamic_1978, keogh_exact_2005}. DTW is used to measure similarity between temporal profiles, and is commonly used in automatic speech recognition to look for similar waveforms when people have different speech cadence. In geophysics, it can also be used as a metric of the agreement of two time series \citep{laperre_dynamic_2020}, without over-penalising for small timing errors \citep{owens_time-window_2018}. It has also been proposed as a way to project solar-wind stream fronts from observing spacecraft at the L1 point to the nose of the Earth's magnetopshere \citep{prchlik_multi-spacecraft_2018}. Here, DTW will be used to assess the connection lines (or ``warp paths'') that connect similar features in two time series. For this purpose, we use the \citet{meert_wannesmdtaidistance_2020} implementation in python.

In brief, the DTW can be summarised as follows: For two data sequences, $A$ ($a_1, a_2, a_3 ... a_M$) and $B$ ($b_1, b_2, b_3 ... b_N$), we seek to match every index in $A$ with one or more indices in $B$ in a way that minimises the difference between the sequences. This matching is subject to a number of rules (here written only for sequence $A$, though they also apply to $B$):
\begin{itemize}
    \item Every index in $A$ must be matched with one or more indices in $B$ 
    \item $a_1$ must map to $b_1$, though it may also map to larger indices in $B$ 
    \item $a_M$ must map to $b_N$, though it may also map to smaller indices in $B$ 
    \item Mapping must be monotonically increasing. I.e. if $a_i$ maps to $b_j$, then $a_{i+1}$ must map to either $b_j$ and/or $b_{j+1}$ (and possibly also higher $B$ indices).
\end{itemize}

A cost function, based on the difference between the mapped indices, is then minimised to give the optimal match. Additional constraints can also be used. It is common to impose a locality constraint, such that mappings between $a_i$ and $b_j$ must meet the criterion $|i-j| < w$, where $w$ is the maximum window length over which mapping is permitted.

An example of DTW applied to two $V$ time series is shown in Figure \ref{fig:dtwmatrix}. Panels a and b show $V$ at Earth produced by the linearly evolving solar-wind model for two consecutive Carrington rotations, $N$ and $N+1$, respectively. Figure \ref{fig:dtwmatrix}c shows the cumulative Euclidean distance between the $V$ values when different sections of the time series are aligned. This alignment is not a fixed time offset, but a variable ``warping'' of the time series. If there was no time evolution, the optimum path (shown in black) would lie along the $y=x$ line (shown in white). During long intervals of slow wind, such as days 20 to 25, there are no features for the algorithm to match and a low cumulative distance can be achieved for a large range of different warpings. At the slow/fast wind interface, however, such as around days 2 and 14, there is a very clear optimum warping, in both cases the structure in CR $N$ is encountered earlier in CR $N+1$. Thus the sharp gradients in $V$ act as `anchor points' for the DTW scheme. As discussed further in Section \ref{sec:discussion}, at times when there is little structure in $V$, it may be necessary to use additional solar wind parameters to provide the required anchor points and produce accurate warp paths between the two time series.

The same example time series are shown in Figure \ref{fig:dtw_explanation} as a function of Carrington longitude and time, with $V$ as the colour scale. The black lines show identified optimal warp paths between the $V$ time series. As the fast solar wind streams are moving to increasing Carrington longitude between the two Carrington rotations, most of these connection lines have a positive gradient. At the point of interest, shown as a red dot, the closest connector is shaded red. It connects fast solar wind at both ends and thus (correctly) predicts fast wind at the reconstruction point. However, the exact magnitudes of $V$ at the start and end of the best warp path are different, so we take a weighted average:

\begin{eqnarray}
    V_{DTW} (t_0, \phi_0) &=& \frac{t_{+W} - t_0}{t_{+W} - t_{-W}} V (t_{-W},\phi_{-W}) \nonumber \\
    & &  +   \frac{t_0 - t_{-W}}{t_{+W} - t_{-W}} V (t_{+W},\phi_{+W})
\end{eqnarray}
where $t_{-W}$ and  $t_{+W}$ are the times of the start and end of the optimal warp paths, respectively, and $\phi_{-W}$ and $\phi_{+W}$ are the associated Carrington longitudes.

For each time and longitude at which $V$ is to be reconstructed, we produce two 27-day time series at 1-hour resolution from the synthetic near-Earth $V$, centred on $t_+$ and $t_-$. We apply DTW to these two time series, allowing 5 days at the start end of the time series to be ignored if it produces a lower average distance, so as to minimise edge effects. We also set a maximum warp window of 5 days, to avoid connecting different solar wind streams. This means that DTW will not be able to recognise very rapid solar wind evolution. The results presented here are insensitive to the choice of these parameters. Optimal smoothing of the input time series and choice of the edge- and warp-windows for use with real in-situ data will be invested in a future paper in this series.

%================================================================================
\section{Results}
\label{sec:results}
%================================================================================

\begin{figure}
	\includegraphics[width=\columnwidth]{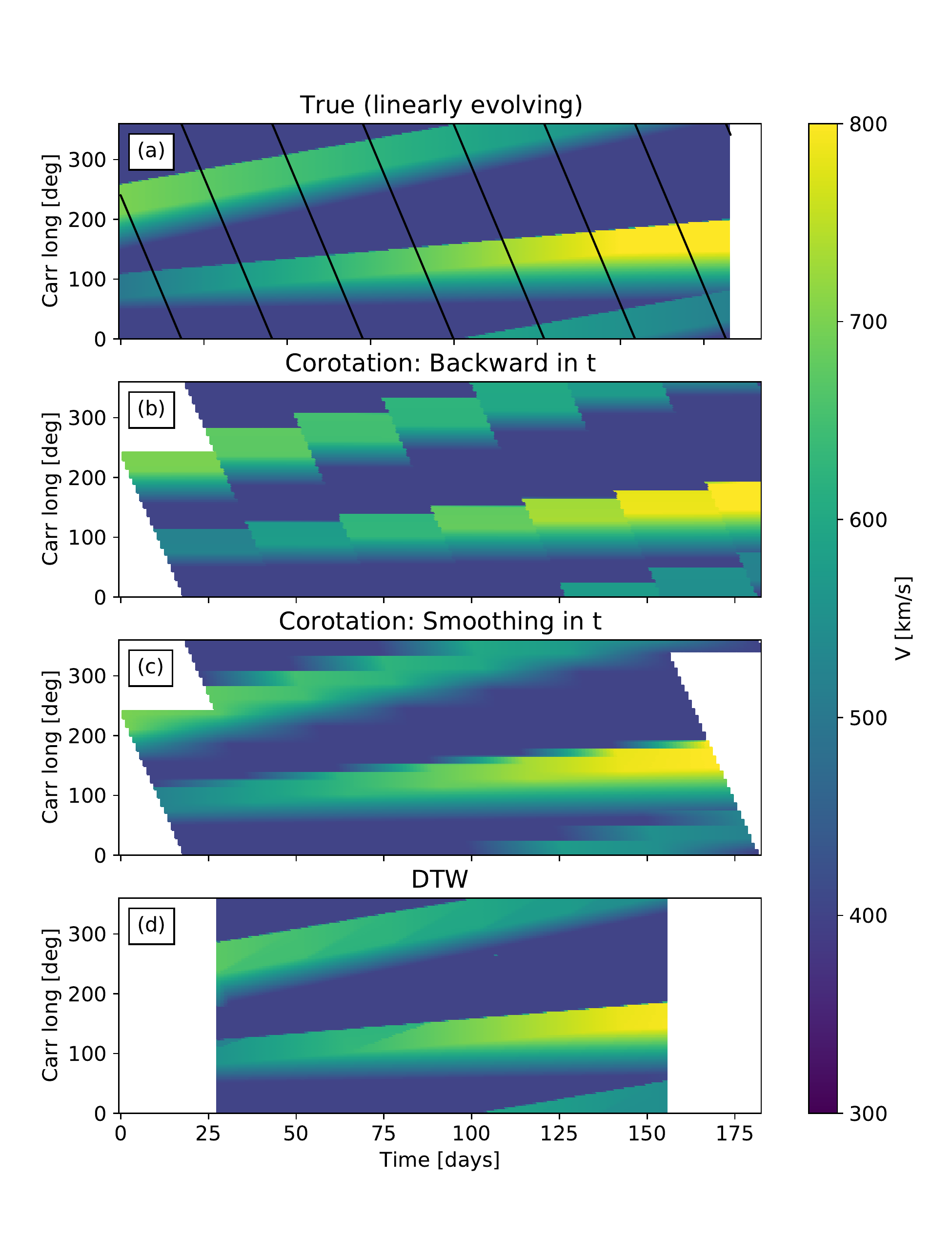}
    \caption{Solar wind speed as a function of longitude and time. (a) The model of linearly evolving solar wind structure, from which synthetic $V$ time series are produced at Earth orbit (black lines). The bottom three panels show  $V$ reconstructed from synthetic near-Earth observations by three methods. These are (b) corotation backwards in time, (c) corotation smoothed backwards and forwards in time, and (d) dynamic time warping.}
    \label{fig:models_vlin}
\end{figure}

\begin{figure}
	\includegraphics[width=\columnwidth]{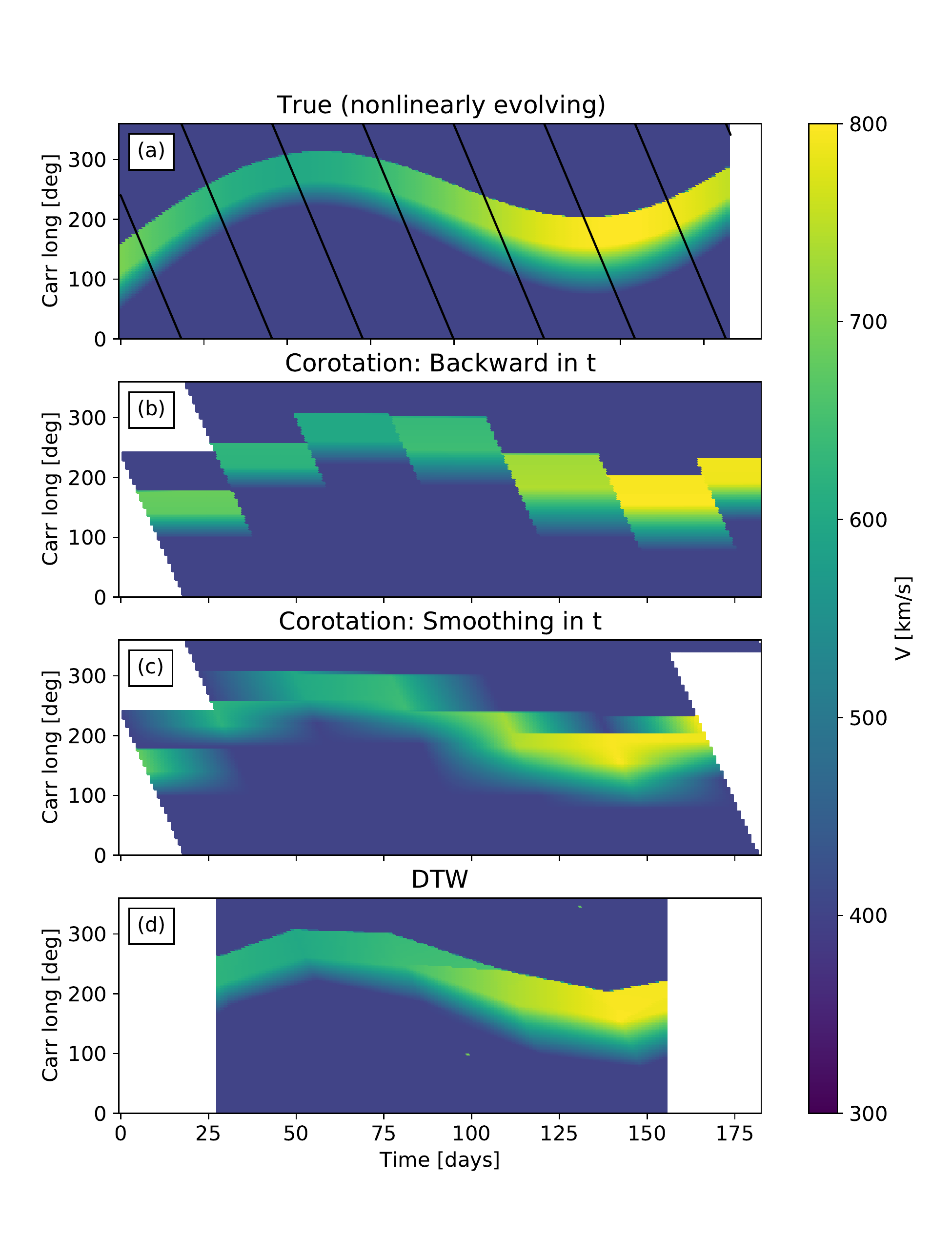}
    \caption{Solar wind speed as a function of longitude and time. (a) The model of nonlinearly evolving solar wind structure, from which synthetic $V$ time series are produced at Earth orbit (black lines). The bottom three panels show  $V$ reconstructed from synthetic near-Earth observations by three methods. These are (b) corotation backwards in time, (c) corotation smoothed backwards and forwards in time, and (d) dynamic time warping.}
    \label{fig:models_vnonlin}
\end{figure}

Figures \ref{fig:models_vlin} and \ref{fig:models_vnonlin} show solar-wind speed as a function of Carrington longitude and time obtained by applying the three reconstruction techniques to the synthetic near-Earth observations. Reconstructions are made at daily cadence and for 128 longitude bins, though results are unchanged at hourly resolution and 720 longitude bins. 

Figures \ref{fig:models_vlin}b and \ref{fig:models_vnonlin}b clearly show that in the presence of time evolving solar-wind structures corotation (backwards in time) produces discontinuities in the reconstructed solar wind as new observations are introduced. When performing 2D solar wind modelling, this produces unphysical solar wind solutions, as will be demonstrated and quantified in a later paper in this series. By comparing the two fast streams in the linear case (Figure \ref{fig:models_vlin}b), it is also apparent that the issues with corotation are greater for features that are evolving faster in time.

Figures \ref{fig:models_vlin} and \ref{fig:models_vnonlin} show that time-smoothing corotation does indeed reduce discontinuities in $V$ relative to corotation back in time when new observations are introduced. However, there are still visible artefacts, particularly for the fast evolving solar wind streams.

The result of the DTW reconstruction method applied to the linearly evolving solar wind is shown in Figure \ref{fig:models_vlin}d. The agreement with the true state (Figure \ref{fig:models_vlin}a) appears, by eye, to be very good. In particular, there are no apparent artefacts at the observation locations, which are obvious for both the simple corotation back in time and the time-smoothing corotation methods. 

\begin{figure}
	\includegraphics[width=\columnwidth]{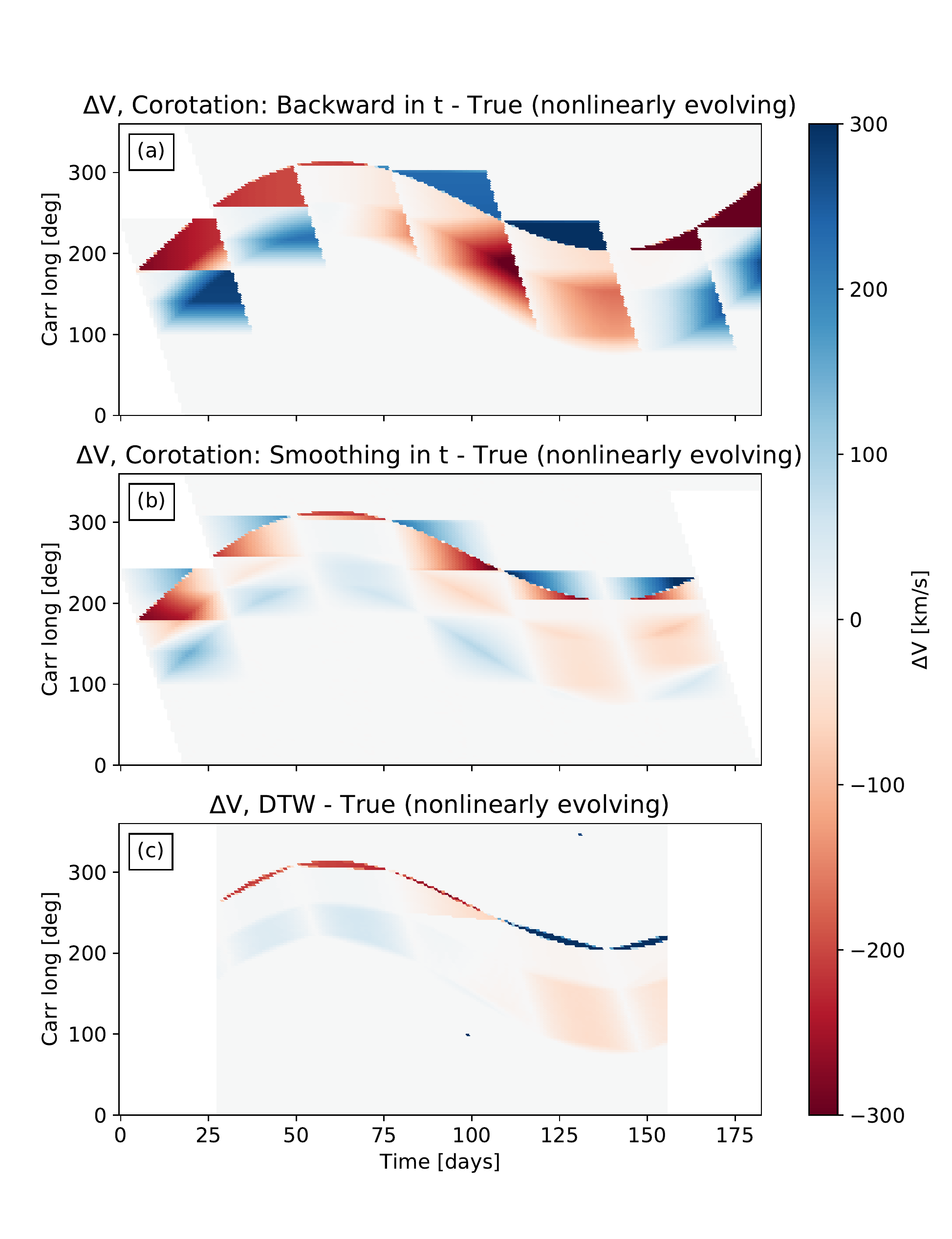}
    \caption{The $V$ error between the reconstruction methods and the true state of the nonlinearly evolving solar wind, as a function of Carrington longitude and time. (a) Corotation backwards in time. (b) Corotation smoothed backwards and forwards in time. (c) Dynamic time warping.}
    \label{fig:models_dvnonlin}
\end{figure}

For the nonlinear solar wind evolution, shown in Figure \ref{fig:models_dvnonlin}, DTW again provides a better visual match than either of the corotation methods. However, unlike the linear case, the location of observations can be seen in the DTW reconstruction as a change in gradient in the longitude of the fast stream with time.

\begin{figure}
	\includegraphics[width=\columnwidth]{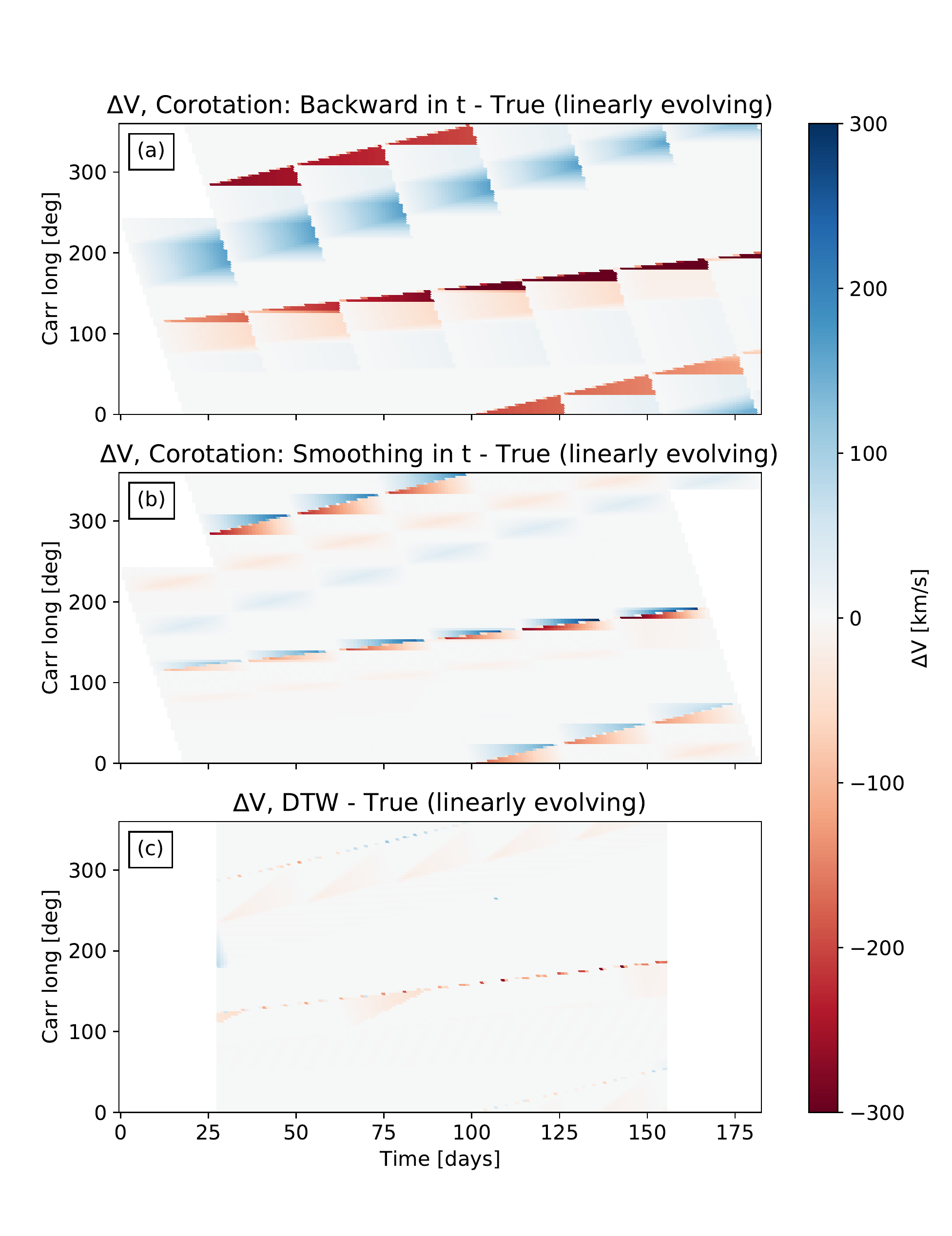}
    \caption{The $V$ error between the reconstruction methods and the true state of the linearly evolving solar wind, as a function of Carrington longitude and time. (a) Corotation backwards in time. (b) Corotation smoothed backwards and forwards in time. (c) Dynamic time warping.}
    \label{fig:models_dvlin}
\end{figure}

\begin{table}
	\centering
	\caption{The mean absolute error, MAE, in $V$ between the true state and reconstructed state from synthetic near-Earth time series. Values are computed between days 27 and 157, when all methods are valid.}
	\label{tab:mae}
	\begin{tabular}{lcc} % four columns, alignment for each
		\hline
		  & <|$\Delta$V|>  (linear) & <|$\Delta$V|>  (nonlinear) \\
		\hline
     	Corotation (backward in t) & 28.5 km s$^{-1}$ & 35.0 km s$^{-1}$ \\
		Corotation (smoothed in t) & 11.0 km s$^{-1}$ & 17.1 km s$^{-1}$ \\
		Dynamic time warping       & 1.8 km s$^{-1}$ & 8.3 km s$^{-1}$ \\
		\hline
	\end{tabular}
\end{table}

In order to visualise these differences more clearly, we also compute $dV$, the difference between the true state and the reconstruction. The linearly-evolving solar wind is shown in Figure \ref{fig:models_dvlin}. The large positive and negative $dV$ values for the two corotation methods (Figures \ref{fig:models_dvlin}a and b, respectively) are almost completely removed by DTW (Figure \ref{fig:models_dvlin}c). The average values over the whole domain, between times of 27 and 157 days, are summarised Table \ref{tab:mae}. The time-smoothing corotation produces average errors about a factor of 2.5 lower than backwards-in-time corotation. DTW provides nearly an order of magnitude further reduction in average errors from time-smoothing corotation.

Figure \ref{fig:models_dvnonlin} shows $dV$ for the nonlinearly evolving solar wind. For DTW, the errors are largely confined close to the high-speed stream front, whereas they are spread over a greater longitude range for the corotation methods. Table \ref{tab:mae} shows that the time-smoothing corotation produces average errors about a factor of two lower than backwards-in-time corotation, with DTW providing a further factor of two reduction from time-smoothing corotation.

\begin{figure}
	\includegraphics[width=\columnwidth]{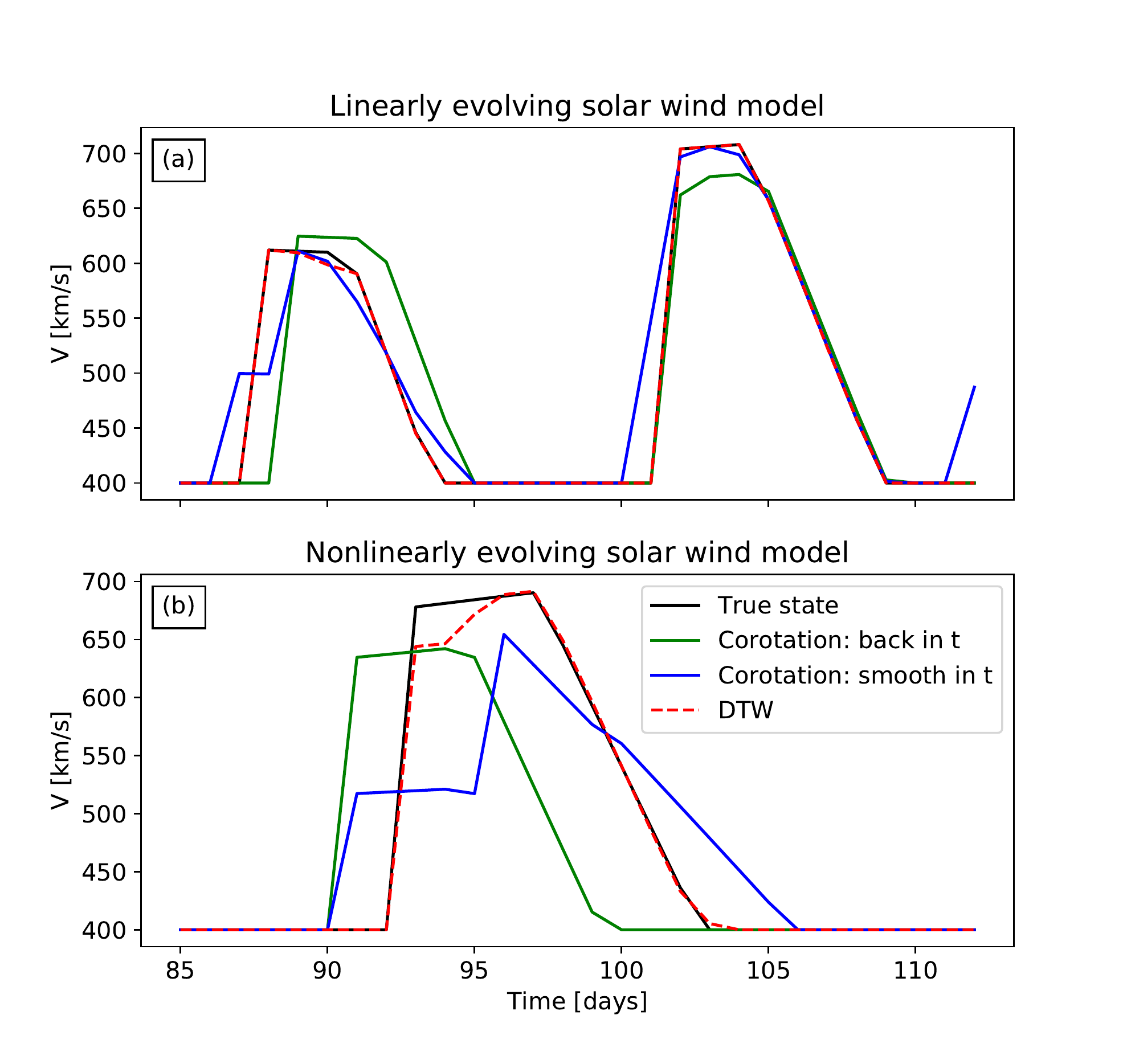}
    \caption{Example of $V$ time series from a point 180$^\circ$ ahead of Earth in its orbit (i.e. the furthest point from an available observation). (a) The linearly evolving solar-wind model. (b) The nonlinearly evolving solar-wind model. Black lines show the true state. Coloured lines show the reconstructed values using the near-Earth $V$ time series using (green) corotation back in time, (blue) corotation smoothed back and forwards in time, and (red dashed line) dynamic time warping.}
    \label{fig:models_examplets}
\end{figure}

It is instructive to look in more detail at the reconstructed $V$ time series at locations distant from observations. Figure \ref{fig:models_examplets} shows time series reconstructed 180$^\circ$ ahead of Earth in its orbit (i.e., at the longitude furthest from available observations). For both the linear and nonlinear cases, corotation back in time (green lines) reproduces the correct general form of the solar wind structure, but  with systematic timing and/or magnitude offsets in the high-speed streams.

The systematic timing errors are also generally removed by time smoothing the corotation reconstruction (blue lines). However the time profile of the high-speed stream has been altered. For the nonlinearly evolving solar wind shown in \ref{fig:models_examplets}b, the true state (black line) shows a single rise in $V$ at day 92.5, whereas time smoothing produces both a jump to the left (at day 90) and a step to the right (at day 96). This results from different $V$ values obtained from averaging the backwards and forwards in time corotations to give intermediate-speed wind, matching neither observation. This results in the two-step $V$ time series profile for the high-speed streams. In a later paper, the effect of this on the reconstructed solar wind at the outer planets will be quantified.

The timing, magnitude and wave-form of the DTW reconstructed fast streams (red dashed lines) is in better agreement with the true state than the corotation methods. For the nonlinearly evolving solar wind (Figure \ref{fig:models_examplets}b), it can be seen that the $dV$ at the leading edge of the fast stream is actually a small magnitude error, rather than the timing errors present for the corotation methods.

\begin{figure}
	\includegraphics[width=\columnwidth]{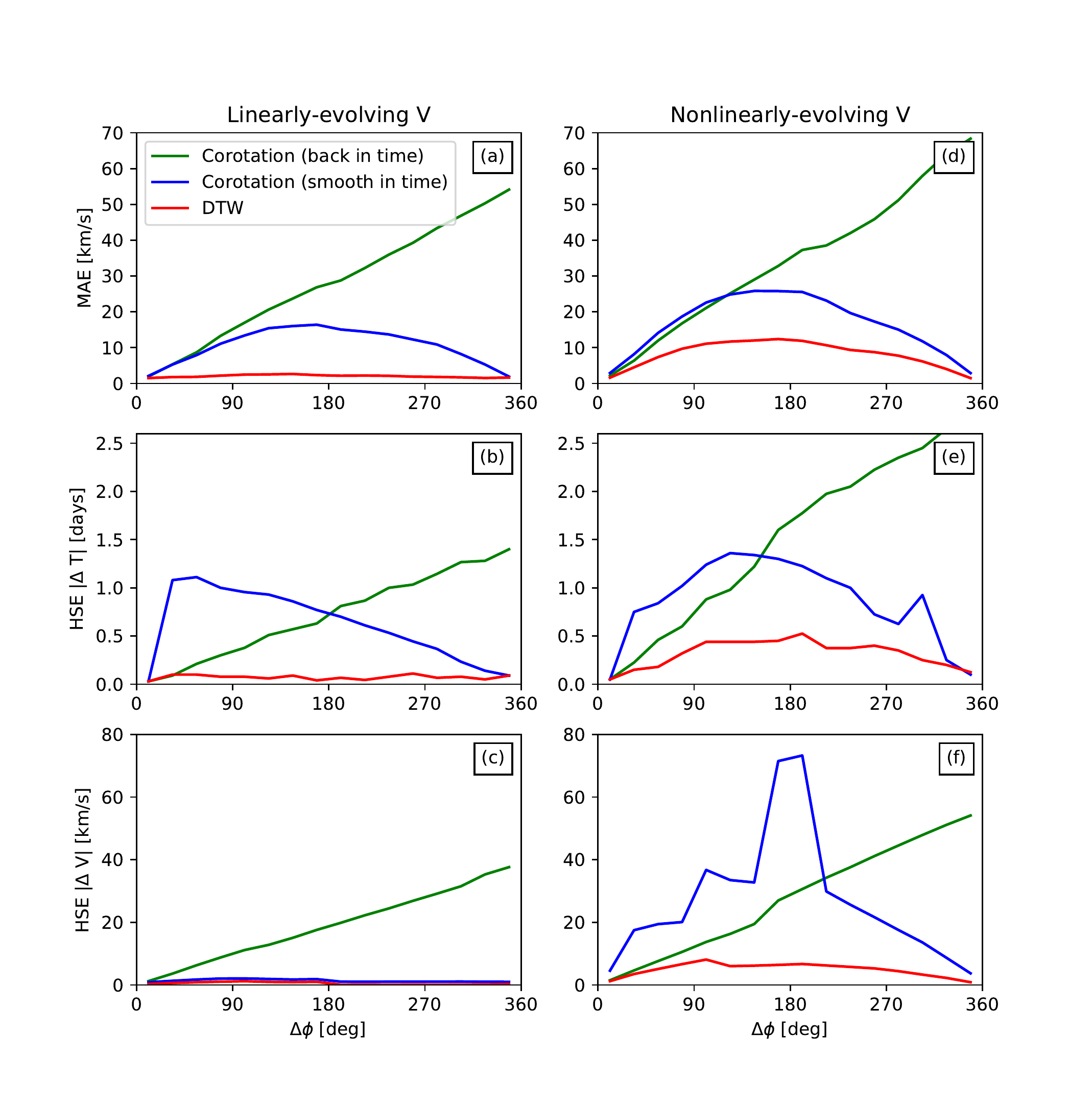}
    \caption{Metrics for the three reconstruction methods at different longitudes relative to Earth, $\Delta \phi$. Left- and right-hand panels show the linearly- and nonlinearly-evolving solar wind, respectively. (a) and (b): The mean-absolute error in $V$. (c) and (d): The timing error, $\Delta T$, in the high-speed streams. (e) and (f): The magnitude error, $\Delta V$, in the high-speed streams. Green lines show corotation back in time, blue show corotation smoothed in time, red show DTW.}
    \label{fig:metrics}
\end{figure}

Figure \ref{fig:models_examplets} only considers a single location relative to Earth. In order to look more generally at the performance of the reconstruction methods, we produce time series for the true state and reconstructions at a range of longitudes relative to Earth, $\Delta \phi$. These are summarised in Figure \ref{fig:metrics}.

For each time series, we compute the mean absolute error (MAE) in $V$, shown in Figures \ref{fig:metrics}a and \ref{fig:metrics}b for the linearly and nonlinearly evolving solar wind, respectively. For corotation back in time (green lines), the MAE increases almost linearly with $\Delta \phi$, and produces a large discontinuity at the $\Delta \phi$ = 0/360$^\circ$ position. Corotation smoothed in time removes this discontinuity, with MAE = 0 at $\Delta \phi$ = 0$^\circ$ and $\Delta \phi$ = 360$^\circ$. MAE is maximised at $\Delta \phi$ = 180$^\circ$. MAE for DTW (red lines) exhibits the same basic behaviour, but with reduced amplitude.

We also apply high-speed enhancement (HSE) analysis to each time series \citep{owens_event-based_2005,macneice_validation_2009,macneice_assessing_2018}. This event-based analysis identifies a HSE using a minimum speed gradient, then associates HSEs between the observed and reconstructed time series. This allows us to determine the timing and magntiude errors in individual high-speed streams. Figures \ref{fig:metrics}c and \ref{fig:metrics}d show that corotation back in time introduces timing errors, $\Delta T$, which grow linearly with $\Delta \phi$. For the non-linearly evolving solar wind, corotation smoothed in time shows similar behaviour to MAE. For the linear case, however, $\Delta T$ peaks at low $\Delta \phi$, at a value larger than corotation back in time, then declines with $\Delta \phi$. This is due to the two-step $V$ profile, with the HSE analysis jumping between defining the event start between the first and second $V$ increase. For DTW, $\Delta T$ is effectively zero for the linear case, but peaks at $\Delta \phi$ = 180$^\circ$ in the nonlinear case, though again at lower values than either of the corotation methods. The HSE magnitude errors, $\Delta V$, are shown in Figures \ref{fig:metrics}e and \ref{fig:metrics}f. Of note is the $\Delta V$ for corotation smoothed in time in the nonlinearly evolving solar wind case (i.e., the blue line in Figure \ref{fig:metrics}f). There is a sharp peak in $\Delta V$ around $\Delta \phi$ = 180$^\circ$. This is again due to the two-step waveform that time smoothing introduces, with the HSE identifying the first front as the HSE and producing a low maximum speed.

\section{Conclusions}
\label{sec:discussion}

Observations of planetary magnetospheres benefit from local solar wind context. For the outer planets, this is typically provided using solar wind models initiated from in situ observations around 1 AU. This is Paper 1 in a series which investigates and validates methods for reconstructing solar wind speed, $V$, at all longitudes from single-point in-situ observations. The standard method is corotation, in which $V$ at a given Carrington longitude is equal to the last available observation at that longitude. Thus corotation-based reconstruction assumes no time variation in the solar wind structure. In order to quantify the effect of this, and explore a technique that relaxes the time-stationary assumption, we constructed two models of solar wind evolution and produce synthetic in-situ observations from which the global structure is to be reconstructed; One model evolves the solar wind structure linearly with time, and one model wherein the solar wind structure evolves nonlinearly. 

Using these models, time smoothing -- in which the reconstructed speed at a given point is a linear combination of the corotated speed both forwards and backwards in time -- is shown to greatly improve $V$ reconstruction over the standard corotation back in time. In particular, the discontinuity in reconstructed $V$ produced by the introduction of new observations is greatly reduced by the use of time smoothing. However, there are clearly still issues, particularly in reconstructing the nonlinearly evolving solar wind. In general, corotation with time smoothing removes systematic offsets in the timing of fast streams, but changes the waveform of reconstructed fast streams, spreading them over a longer time interval and reducing the peak amplitude. 

To address these issues we adapt a data analysis technique employed in automatic speech recognition to account for different speech cadence. Dynamic time warping (DTW) computes the agreement between temporal profiles when connecting different points within the profile. By finding the `warp path' which maximises the agreement, we are able to interpolate between the same solar wind structures, rather than solar wind at the same Carrington longitudes. Applying the DTW method to the linearly evolving solar wind produces an almost perfect reconstruction, with an order-of-magnitude reduction in $V$ errors even compared to corotation time smoothing. For the nonlinear case, the improvements were not as large, but still significant over the corotation methods. The implications for using these boundary conditions for solar-wind prediction at the outer planets will be investigated using the HUXt solar wind model \citep{owens_computationally_2020} in future work, though the results are expected to be generally applicable to any model using the same input boundary conditions.

We note that real in-situ observations can contain transient structures in the form of interplanetary coronal mass ejections (ICMEs). Assuming ICMEs repeat in subsequent Carrington rotations, or are seen at spacecraft which are well separated in longitude, will create major errors in corotation-based schemes and will produce incorrect warping paths in the DTW approach. Methods to deal with transient features will be investigated in a subsequent paper in the series.

For simplicity, this study has only looked at solar-wind speed and synthetic observations from a single spacecraft. The same methodology can be applied to multiple spacecraft observations near 1 AU, as will later be demonstrasted using near-Earth and STEREO observations. 

The warping paths determined from $V$ observations can be applied to other solar wind parameters, allowing a full set of input conditions for magnetohydrodynamic models \citep[e.g.][]{zieger_statistical_2008,tao_magnetic_2005}. (Note that it would not be desirable to determine and apply different warp paths for different solar-wind parameters, as this result in unphysical solutions. E.g. potentially moving a period of southward heliospheric magnetic field from a region of fast to slow solar wind.) We also note that it is possible to determine a single set of warp paths from the matching of multiple time series. Thus, it is possible to find a single set of warp paths that provide the best match to multiple solar wind parameters, such as speed, magnetic field polarity and density. This will be beneficial if the DTW method is under constrained by insufficient common ``anchor points'' within a single parameter time series. For the synthetic time series considered in this study, this was not an issue. Unambiguous testing of the method with real solar-wind observations is difficult as the true state is not known. However, in the final paper in the series we will determine the robustness of the warp paths as new information is provided. For real solar wind observations, determining the most accurate warping paths may require pre-processing of the data, such as smoothing high-frequency `noise'. This will also be investigated.

\section*{Acknowledgements}

Work was part-funded by Science and Technology Facilities Council (STFC) grant numbers ST/R000921/1 and ST/V000497/1, and Natural Environment Research Council (NERC) grant numbers NE/S010033/1 and NE/P016928/1. 

%%%%%%%%%%%%%%%%%%%%%%%%%%%%%%%%%%%%%%%%%%%%%%%%%%
\section*{Data Availability}

No data was used in this study. All model and analysis code is available in the python language from \url{www.github.com/University-of-Reading-Space-Science/SolarWindInputs_DTW}.

%%%%%%%%%%%%%%%%%%%% REFERENCES %%%%%%%%%%%%%%%%%%

% The best way to enter references is to use BibTeX:

% Alternatively you could enter them by hand, like this:
% This method is tedious and prone to error if you have lots of references
%\begin{thebibliography}{99}
%\bibitem[\protect\citeauthoryear{Author}{2012}]{Author2012}
%Author A.~N., 2013, Journal of Improbable Astronomy, 1, 1
%\bibitem[\protect\citeauthoryear{Others}{2013}]{Others2013}
%Others S., 2012, Journal of Interesting Stuff, 17, 198
%\end{thebibliography}

%%%%%%%%%%%%%%%%%%%%%%%%%%%%%%%%%%%%%%%%%%%%%%%%%%

%%%%%%%%%%%%%%%%% APPENDICES %%%%%%%%%%%%%%%%%%%%%

%%%%%%%%%%%%%%%%%%%%%%%%%%%%%%%%%%%%%%%%%%%%%%%%%%

% Don't change these lines
\bsp	% typesetting comment
\label{lastpage}
\end{document}